\documentclass[12pt,showpacs,prd]{revtex4}
\usepackage{epsfig,amssymb,amsfonts}
\begin{document}
\title{Chiral symmetry breaking solutions for QCD in the truncated Coulomb gauge}
\author{P. J. A. Bicudo}
\affiliation{Grupo Te\'orico de Altas Energias (GTAE), Centro de F\'\i sica das
Interac\c c\~oes Fundamentais (CFIF),
Departamento de F\'\i sica, Instituto Superior T\'ecnico, Av. Rovisco Pais, P-1049-001 Lisboa, Portugal}
\author{A. V. Nefediev}
\affiliation{Institute of Theoretical and Experimental Physics, 117218,\\ B.Cheremushkinskaya 25,
Moscow, Russia}
\newcommand{\be}{\begin{equation}}
\newcommand{\bea}{\begin{eqnarray}}
\newcommand{\ee}{\end{equation}}
\newcommand{\eea}{\end{eqnarray}}
\newcommand{\ds}{\displaystyle}
\newcommand{\low}[1]{\raisebox{-1mm}{$#1$}}
\newcommand{\loww}[1]{\raisebox{-1.5mm}{$#1$}}
\newcommand{\lmn}{\mathop{\sim}\limits_{n\gg 1}}
\newcommand{\vpint}{\int\makebox[0mm][r]{\bf --\hspace*{0.13cm}}}
\newcommand{\tooo}{\mathop{\to}\limits_{m\to 0}}
\newcommand{\vp}{\varphi}

\begin{abstract}
In this paper we study the power-like confining potentials $r^\alpha$. The region of
allowed $\alpha$'s is identified, the mass-gap equation is constructed for an arbitrary
$\alpha$ and solved for several values of the latter, and the vacuum energy and the
chiral condensate are calculated. The question of replica
solutions to the mass-gap equation for such potentials is addressed and it is demonstrated
that the number of replicas is infinite for any $\alpha$, as a consequence of the peculiar
behaviour of the quark self-energy in the infrared domain.
\end{abstract}
\maketitle

\section{Introduction}

The problem of the spontaneous breaking of chiral symmetry (SBCS) and its relation to confinement is one of
the QCD cornerstones. Although the basic ideas of SBCS are already a textbook topic, this problem still lies at
the crossroad of many studies and approaches to QCD. In this paper we exploit the
potential model for QCD,
whose origins can be traced back to QCD in the truncated Coulomb gauge and which is proven to be successful
in studies of the low-energy phenomena in QCD (see, for example, \cite{pipi}). This class of models can be indicated as
Nambu-Jona-Lasinio-(NJL-)type models \cite{NJL} with the current-current quark interaction and the corresponding
formfactor coming from the bilocal gluonic correlator. The standard approximation in such type of models is
to neglect the retardation and to approximate the gluonic correlator by a confining potential of a certain
form. The power-like potentials, which are the most natural candidates for the role of the confining force, are
the subject of the present investigation. In the course of this paper we
\begin{itemize}
\item revisit the problem of SBCS for the for power-like confining potentials
$V(r)=K_0^{1+\alpha}r^{\alpha}$ with $\alpha\geqslant 0$;
\item restrict the range of allowed $\alpha$'s and, for several values of the latter, find numerically
solutions to the corresponding mass-gap equation, as well as the vacuum energy density, and the chiral
condensate for the chirally noninvariant vacuum of the theory;
\item study in detail the problem of existence of the replica solutions to the mass-gap
equation for power-like confining potentials.
\end{itemize}

The problem of instability of the chirally invariant vacuum for power-like confining potentials was studied
in detail in the middle of 80's by the Orsay group \cite{Orsay} and such an instability was proved for the
range $0\leqslant\alpha<3$. For numerical studies the harmonic oscillator type potential, $\alpha=2$, was chosen by
these authors, as well as by the Lisbon group \cite{Lisbon}, and a set of results for hadronic
properties were obtained in the framework of the given model. In this paper we study the mass-gap equation
for an arbitrary value of $\alpha$ ranging from 0 to 2, with special attention payed to $\alpha$'s
close to unity, since the linearly rising potential is known to be preferred by phenomenology as the most
successful candidate for the confining force giving the correct Regge trajectories behaviour, possessing a
clear connection to the QCD string, and so on (see \cite{linear} and references therein). It is also claimed
to be singled out by the lattice
calculations. We exclude the region $\alpha>2$ since the corresponding
mass-gap equation diverges for such $\alpha$'s. On the other hand, it would be hard to justify the use of such a
strong confining force in phenomenological models for QCD.
We solve the mass-gap equation explicitly for several values of $\alpha$ from the allowed region
and demonstrate that the chiral angle, the vacuum energy density, and the chiral condensate
are smooth slow functions of the form of the confining potential, so that
the results obtained for the potential of a given form --- the linear confinement being the most justified
and phenomenologically successful choice --- have a universal nature for any quark-quark kernel of such
a type.

Following the set of recent publications devoted to possible multiple solutions for the chirally noninvariant
vacuum in QCD \cite{replica1,replica2} (see also \cite{replica3} where a similar conclusion was made in a different
approach), we address the question of replicas existence for various power laws $r^\alpha$. We find that for
a whole range of allowed powers, $0\leqslant\alpha\leqslant 2$, replica solutions do
exist, similarly to the
case of $\alpha=2$ studied in detail in \cite{Orsay,Lisbon}. We give the profiles of
several replicas for the linear confinement and
argue that the number of such solutions is infinite for any power $\alpha$, including the weakest,
logarithmic, potential which corresponds to $\alpha=0$. We argue that the source of replicas
is the infrared behaviour of the single-quark self-energy --- the dressed quark dispersive law $E(p)$ ---
which, for small values of the quark momentum $p$, becomes a negative sharp function of
$p$ thus
enabling fast oscillations of the chiral angle with the frequency increasing with vanishing momentum $p$.
Since this property of the quark dispersive law is expected to be an integral part of any confining
interaction, then we confirm the conclusion made in \cite{replica2} that
\lq\lq$\ldots$ across all these different quark kernels, the existence of vacuum replicas should constitute
the rule rather than the exception." We argue that, in real QCD, with the confining
interaction flattening at large distances due to the effect of the QCD string breaking,
the number of replicas becomes finite. We find the
parameter of the SBCS given by the replica solutions to decrease fast with the number of nodes of the chiral
angle so that one has a well defined perturbative series in replicas and, therefore,
the account only for the first replica may be sufficient in many phenomenological applications (the details of
the formalism which allows one to incorporate replicas into quark models can be found in
ref.~\cite{replica2}).

The paper is organised as follows: In the second section we give the necessary details of
the formalism and derive the mass-gap equation for the power-like confining potential,
which is studied in detail in the third section, first, qualitatively, then quantitatively,
and, finally, numerically. The mass-gap equation is solved numerically for several
values of $\alpha$ and the chiral condensate and the excess of the vacuum energy
density over the trivial solution are calculated for the found solutions. In the fourth
section, devoted to replicas, we demonstrate how an infinite number of solutions to the
mass-gap equation appears and build explicitly a couple of replicas for the linear confinement. Our
conclusions are the subject of the last, fifth, section.

\section{The mass-gap equation}

The chiral model which we use for our studies is given by the Hamiltonian with the current-current
interaction parametrised by the bilocal correlator $K_{\mu\nu}^{ab}$,
\be
\label{H}
H=\int d^3 x\bar{\psi}(\vec{x},t)\left(-i\vec{\gamma}\cdot\vec{\bigtriangledown}\right)\psi(\vec{x},t)+
\frac12\int d^3
xd^3y\;J^a_\mu(\vec{x},t)K^{ab}_{\mu\nu}(\vec{x}-\vec{y})J^b_\nu(\vec{y},t),
\ee
where the quark current is
$J_{\mu}^a(\vec{x},t)=\bar{\psi}(\vec{x},t)\gamma_\mu\frac{\lambda^a}{2}\psi(\vec{x},t)$, and
the gluonic correlator is approximated by a potential,
\be
K^{ab}_{\mu\nu}(\vec{x}-\vec{y})=g_{\mu 0}g_{\nu 0}\delta^{ab}V_0(|\vec{x}-\vec{y}|),
\ee
with
\be
V_0(|\vec{x}|)=K_0^{\alpha+1}|\vec{x}|^{\alpha}.
\label{potential}
\ee

In order to include the logarithmic potential into
consideration the obvious modification of the
potential is needed:
\be
\left.V_0(|\vec{x}|)\to {\tilde V}_0(|\vec{x}|)=K_0
\frac{(K_0|\vec{x}|)^\alpha-1}{\alpha}
\right|_{\alpha\to 0}=K_0\ln(K_0|\vec{x}|).
\label{ln}
\ee

The model contains the only dimensional parameter --- the strength of the confining force $K_0$.
For further convenience we shall
consider a modified version of the potential (\ref{potential}) \cite{Orsay},
\be
V_0(|\vec{x}|)=K_0^{\alpha+1}|\vec{x}|^{\alpha}e^{-m|\vec{x}|},
\label{potential2}
\ee
where $m$ plays the role of the regulator for the infrared behaviour of the interaction. The limit $m\to 0$ is
understood.

The standard technique used in such models is the
Bogoliubov-Valatin transformation from bare to dressed
quarks parametrised by the chiral angle --- the main entity defining the chiral symmetry breaking, the
structure of the BCS vacuum of the theory, as well as the properties of the hadronic states built over this
vacuum \cite{Orsay,Lisbon}. For the application
of this technique to the two-dimensional QCD
\cite{tHooft} see the papers
\cite{2d}. We choose the following parameterisation:
\be
\psi(\vec{x},t)=\sum_{\xi=\uparrow,\downarrow}\int\frac{d^3p}{(2\pi)^3}e^{i\vec{p}\vec{x}}
[b_{\xi}(\vec{p},t)u_\xi(\vec{p})+d_{\xi}^\dagger(-\vec{p},t)v_\xi(-\vec{p})],
\label{psi}
\ee
\be
\begin{array}{rcl}
u(\vec{p})&=&\frac{1}{\sqrt{2}}\left[\sqrt{1+\sin\vp_p}+
\sqrt{1-\sin\vp_p}\;(\vec{\alpha}\hat{\vec{p}})\right]u(0),\\
v(-\vec{p})&=&\frac{1}{\sqrt{2}}\left[\sqrt{1+\sin\vp_p}-
\sqrt{1-\sin\vp_p}\;(\vec{\alpha}\hat{\vec{p}})\right]v(0),
\end{array}
\label{uandv}
\ee
\be
b_{\xi}(\vec{p},t)=e^{iE_pt}b_{\xi}(\vec{p},0),\quad
d_{\xi}(-\vec{p},t)=e^{iE_pt}d_{\xi}(-\vec{p},0),
\label{bandd}
\ee
where $E_p$ (the shorthand notation for $E(p)$) stands for the dispersive law of the
dressed quarks, the chiral angle $\vp(p)$
(we also use the shorthand notation $\vp_p$ for it) varies in the
range $-\pi/2<\vp_p\leqslant \pi/2$
with the boundary conditions $\vp(0)=\pi/2$, $\vp(p\to\infty)\to 0$.

The Hamiltonian (\ref{H}) normally arranged in the basis (\ref{bandd}) splits into the vacuum energy, the
quadratic and the quartic parts in terms of the quark creation/annihilation operators. For the vacuum energy
density one has
\be
{\cal E}_{\rm vac}[\varphi]=\frac{1}{V}\langle 0| TH[\varphi]|0\rangle=-\frac{g}{2}\int\frac{d^3p}{(2\pi)^3}
\biggl(A(p)\sin\vp_p+[B(p)+p]\cos\vp_p\biggr),
\label{Evac1}
\ee
where $V$ is the three-dimensional volume; the degeneracy factor $g$ counts the number of independent quark degrees of freedom,
\be
g=(2s+1)N_CN_f,
\ee
with $s=\frac12$ being the quark spin; the number of colours, $N_C$, is put to three, and
the number of light flavours, $N_f$, is two. Thus we find that $g=12$. The auxiliary functions $A(p)$
and $B(p)$ are defined as
\be
A(p)=\frac12C_F\int
\frac{d^3k}{(2\pi)^3}V_0(\vec{p}-\vec{k})\sin\vp_k,
\label{A}
\ee
\be
B(p)=p+\frac12C_F\int \frac{d^3k}{(2\pi)^3}\;
(\hat{\vec{p}}\cdot\hat{\vec{k}})V_0(\vec{p}-\vec{k})\cos\vp_k,
\label{B}
\ee
where $C_F=\frac43$ is the $SU(3)_C$ Casimir operator in the fundamental representation.
The actual form of the chiral angle is such that the quadratic part of
the normally ordered Hamiltonian diagonalises, or
alternatively, the vacuum energy takes its minimal value.
The corresponding equation
\be
\frac{\delta {\cal E}_{\rm vac}[\varphi]}{\delta\varphi_p}=0,
\label{massgap}
\ee
known as the mass-gap equation, reads
\be
A(p)\cos\vp_p-B(p)\sin\vp_p=0.
\label{mge}
\ee

For the generalised power-like potential (\ref{potential2}) one can find
\be
A(p)=-C_F\Gamma(\alpha+1)\frac{K_0^{\alpha+1}}{p}\int_{-\infty}^{\infty}\frac{dk}{2\pi}
\frac{k\cos[(\alpha+1)\arctan\frac{k-p}{m}]}{[m^2+(k-p)^2]^{\frac{\alpha+1}{2}}}\sin\vp_k,
\label{A02}
\ee
$$
B(p)=p-C_F\Gamma(\alpha)\frac{K_0^{\alpha+1}}{p^2}\int_{-\infty}^{\infty}\frac{dk}{2\pi}
\left[
\frac{\alpha pk\cos[(\alpha+1)\arctan\frac{k-p}{m}]}{[m^2+(k-p)^2]^{\frac{\alpha+1}{2}}}-
\frac{\cos[(\alpha-1)\arctan\frac{k-p}{m}]}{(\alpha-1)[m^2+(k-p)^2]^{\frac{\alpha-1}{2}}}\right.
$$
\be
\left.-\frac{(k-p)\sin[\alpha\arctan\frac{k-p}{m}]}{[m^2+(k-p)^2]^{\frac{\alpha}{2}}}
\right]\cos\vp_k,
\label{B2}
\ee
where, for the sake of convenience, we continued the integral to the negative values of $k$ assuming
$\cos\vp_{-k}=-\cos\vp_k$, $\sin\vp_{-k}=\sin\vp_k$ (the most natural realisation of these conditions can be
achieved in terms of some even function $m_p$, such that $\sin\vp_p=\frac{m_p}{\sqrt{p^2+m_p^2}}$,
$\cos\vp_p=\frac{p}{\sqrt{p^2+m_p^2}}$, which plays the role of the effective mass of the quark).
Consequently, the
mass-gap equation (\ref{mge}) takes the form:
$$
p^3\sin\varphi_p=C_FK_0^{\alpha+1}\Gamma(\alpha)\int_{-\infty}^{\infty}\frac{dk}{2\pi}\left\{
\frac{\alpha pk\cos[(\alpha+1)\arctan\frac{k-p}{m}]}{[m^2+(k-p)^2]^{\frac{\alpha+1}{2}}}\sin[\vp_p-\vp_k]\right.
$$
\be
+\left.\left(\frac{\cos[(\alpha-1)\arctan\frac{k-p}{m}]}{(\alpha-1)[m^2+(k-p)^2]^{\frac{\alpha-1}{2}}}
+\frac{(k-p)
\sin[\alpha\arctan\frac{k-p}{m}]}{[m^2+(k-p)^2]^{\frac{\alpha}{2}}}\right)\cos\vp_k\sin\vp_p\right\},
\label{mgapg}
\ee
and it is the main object of our studies.

\section{Investigation of the general formula}

\subsection{Qualitative analysis}

As the first step in studies of the general formula (\ref{mgapg}), we perform its simple qualitative analysis.
Using the techniques described in \cite{replica1}, we assume that a solution $\vp^{(0)}(p)$ to this
equation exists, and the vacuum energy is minimal on this solution. If the function $\vp^{(0)}(p/A)$
with an arbitrary scale parameter $0\leqslant A<\infty$ is substituted into the
vacuum energy (\ref{Evac1}), then the function
${\cal E}_{\rm vac}(A)$ must reveal a minimum for $A=1$. Moreover, the corresponding minimum should lie
lower than the one for the trivial solution $\vp(p)\equiv 0$, when there is no dressing of quarks and the chiral
symmetry is unbroken. If the regulator $m$ is removed from the vacuum energy functional, then the strength of
the potential remains the only dimensional parameter in the theory, so that, after a proper rescaling of the
integration variables, one arrives at the simple formula, for an arbitrary number of spatial dimensions $d$,
\be
{\cal E}_{\rm vac}(A)=C_1A^{d+1}+C_2K_0^{\alpha+1}A^{d-\alpha},
\label{Evac11}
\ee
where $C_1$ and $C_2$ are two constants independent of $A$ which are interrelated by the constraint
$\partial{\cal E}_{\rm vac}(A)/\partial A_{|A=1}=0$. The first term in Eq.~(\ref{Evac11}) comes
from the kinetic energy, the second term is due to the interaction.
The following four situations are
possible: i) $0<\alpha<d$, ii) $\alpha>d$, and two boundary cases, iii) $\alpha=d$, and iv) $\alpha=0$.
In the first case the vacuum energy has a double-well form with two minima: trivial for $A=0$ and nontrivial for $A=1$. The difference
${\cal E}_{\rm vac}(A=1)-{\cal E}_{\rm vac}(A=0)$ is negative, so that the chirally nonsymmetric nontrivial
solution is energetically preferable. For the second case one has the interaction term in (\ref{Evac11})
containing negative powers of $A$ and, as a result, the trivial solution, with unbroken
chiral symmetry and which
corresponds to $A=0$, possesses an infinite energy
and therefore it does not exist. In the meantime, a nontrivial solution with $A=1$ still may be present.
The boundary case of $\alpha=d$ leads to the logarithmic dependence of the vacuum energy on the parameter $A$,
\be
{\cal E}_{\rm vac}(A)=C_1A^{d+1}+C_2K_0^{d+1}\ln\frac{A}{K_0},
\label{Evac12}
\ee
so that qualitatively the same conclusion holds --- the theory possesses only chirally nonsymmetric phase.
The two-dimensional QCD \cite{tHooft} is an example of the theory with such a logarithmic dependence
(see \cite{replica1} for the details).

Finally, for the case iv), that is, for the logarithmic potential (\ref{ln})
one has
\be
{\cal E}_{\rm vac}(A)=C_1A^{d+1}+C_2K_0A^d\ln\frac{A}{K_0},
\label{Evac13}
\ee
where the logarithmic growth of the energy, when approaching the trivial solution limit $A=0$, is
cancelled by the power factor $A^d$, so that both, chirally symmetric and nonsymmetric, solutions
coexist in this case, similarly to other potentials with $0<\alpha<d$.

Thus we conclude that once the power of the potential reaches the critical value equal to the number of
spatial dimensions the behaviour of the theory changes drastically, the chirally symmetric phase being swept
off. Meanwhile the qualitative analysis performed above ignored the problem of convergence of the integrals
in the expression for the vacuum energy and in the corresponding mass-gap equation. It also fails to answer
the question as to how many solutions to the mass-gap equation exist. In what follows we
turn to the quantitative and numerical analysis of these problems.

\subsection{Quantitative analysis}

Now we turn to the detailed analysis of the mass-gap equation (\ref{mgapg}), such as the problem of the
convergence, the allowed region for $\alpha$'s, the dependence on the regulator $m$, and so on.

First of all, one can
easily check that the case $\alpha=0$ brings no
difficulties --- the r.h.s. vanishes if the limit $\alpha\to 0$ is taken naively, whereas to arrive at the
mass-gap equation for the logarithmic potential one is to divide the r.h.s. by $\alpha$
(see Eq.~(\ref{ln})) which leads to a finite result after taking the limit $\alpha\to 0$.

For $\alpha=1$ the divergent term proportional to $1/(\alpha-1)$ vanishes on the r.h.s. of Eq.~(\ref{mgapg})
since the cosine of the chiral angle is odd. An accurate expansion of this term for $\alpha\to 1$
brings about logarithmic terms.

Now let us check the largest value of $\alpha$ which does not lead to divergences in the mass-gap equation.
When the regulator $m$ tends to zero, the first term in the curly
brackets in Eq.~(\ref{mgapg}), formally, is the most singular term for $k\sim p$, and it can be written as
\be
\int_{-\infty}^{\infty}\frac{dk}{2\pi}
\frac{pk\cos[(\alpha+1)\arctan\frac{k-p}{m}]}{[m^2+(k-p)^2]^{\frac{\alpha+1}{2}}}\sin[\vp_p-\vp_k].
\ee
In the region $k\sim p$ the integrand admits an expansion in powers $(k-p)^n$:
\be
\frac{\cos[(\alpha+1)\arctan\frac{k-p}{m}]}{2\pi[m^2+(k-p)^2]^{\frac{\alpha+1}{2}}}
\left\{p^2(k-p)\vp'_p+(k-p)^2\left[p\vp'_p+\frac12p^2\vp''_p\right]\ldots\right\},
\ee
where the first term in the curly brackets is odd and, therefore, it vanishes in the integral in
$k$ around $k=p$. The remaining expression, as well as other terms on the r.h.s. of Eq.~(\ref{mgapg}),
behaves as $|k-p|^{1-\alpha}$ and the corresponding
contribution to the mass-gap equation converges for $\alpha<2$, which is the upper limit of the range of valid
powers $\alpha$. For such $\alpha$'s,
\be
\arctan\frac{k-p}{m}\tooo\frac{\pi}{2}{\rm sign}(k-p)+O(m),
\label{arctan}
\ee
and the regulator $m$ can be removed from the mass-gap equation.
For the case of $\alpha=2$ the entire r.h.s. of Eq.~(\ref{mgapg}) vanishes after the substitution
(\ref{arctan}). To be more precise, one has to keep next terms in the expansion (\ref{arctan}),
with positive powers of the regulator $m$, reproducing the well-known representation of the delta function,
$$
\delta(p-k)=\lim_{m\to 0}\frac{1}{\pi}\frac{m}{m^2+(k-p)^2},
$$
and its derivatives. We shall consider this special case separately.

Notice that the proof of the chirally symmetric vacuum instability
given in \cite{Orsay} was based on the consideration of the self-energy functional
$F(\vec{p})=\int\frac{d^3k}{(2\pi)^3}V_0(\vec{p}-\vec{k})(\hat{\vec{p}}\hat{\vec{k}})
\propto\frac{1}{p^\alpha}$, which gives infrared divergent contribution to integrals in $d^3p$ if
$\alpha\geqslant 3$. In the meantime, as demonstrated above, the requirement of finiteness of the mass-gap equation
imposes a stronger restriction on $\alpha$: $\alpha\leqslant 2$.

One encounters no more difficulties for $\alpha's$ within the interval $0\leqslant\alpha<2$ and, when the
regulator $m$ is removed, arrives at the mass-gap equation in the ultimate form:
\begin{eqnarray}
p^3\sin\vp_p&=&C_FK_0^{\alpha+1}\Gamma(\alpha+1)\sin\frac{\pi\alpha}{2}\int_{-\infty}^{\infty}
\frac{dk}{2\pi}\left\{\frac{pk\sin[\vp_k-\vp_p]}{|p-k|^{\alpha+1}}\right.\nonumber\\
&+&\left.
\frac{\cos\vp_k\sin\vp_p}{\alpha-1}\left[\frac{1}{|p-k|^{\alpha-1}}-\frac{1}{\lambda^{\alpha-1}}\right]
\right\}.
\label{mg2}
\end{eqnarray}

We introduced the term $\frac{1}{\lambda^{\alpha-1}}$, with an arbitrary mass
parameter $\lambda$, in order to emphasise the convergence of the integral for
$\alpha=1$. This extra term does not contribute to the integral due to the parity of $\cos\vp_k$.

In particular, the mass-gap equation for the logarithmic potential follows from
(\ref{mg2}) in the limit $\alpha\to 0$, if the proper modification of the potential,
given in Eq.~(\ref{ln}), is applied,
\be
p^3\sin\vp_p=\frac{C_FK_0}{4}\int_{-\infty}^{\infty}\frac{dk}{|p-k|}\left[
pk\sin[\vp_k-\vp_p]-(p-k)^2\cos\vp_k\sin\vp_p\right].
\label{mgln}
\ee

For the case of $\alpha=1$, that is, for the linear confinement,
the formula well known in the literature is readily reproduced (see, for example, \cite{replica1}),
\be
\label{lmge}
p^3\sin\vp_p=C_FK_0^2\int_{-\infty}^{\infty}\frac{dk}{2\pi}\left\{\frac{pk}{(p-k)^2}
\sin[\vp_k-\vp_p]-\ln\frac{|p-k|}{\lambda}\cos\vp_k\sin\vp_p\right\}
\ee
$$
=C_FK_0^2\int_0^{\infty}\frac{dk}{2\pi}\left\{\frac{4p^2k^2}{(p^2-k^2)^2}
\sin[\vp_k-\vp_p]
-\left[\frac{2pk}{(p+k)^2}+\ln\left|\frac{p-k}{p+k}\right|\right]\cos\vp_k\sin\vp_p\right\}.
$$

To complete our investigation let us consider the case of the harmonic oscillator potential, $\alpha=2$. The
Fourier transform of the potential $V(r)=K_0^3 r^2$ is the Laplacian of the three-dimensional delta
function, so that the resulting mass-gap equation becomes differential \cite{Orsay},
\be
p^3\sin\vp_p=\frac12C_FK_0^3\left[p^2\vp''_p+2p\vp_p'+\sin2\vp_p\right].
\label{diffmge}
\ee

Formally, Eqs.~(\ref{mgapg}), (\ref{mg2}) remain valid for $-1<\alpha<0$, and the ultraviolet divergence is
encountered for $\alpha=-1$, that is, for the Coulomb potential. We disregard this region since the resulting
force fails to be confining. Thus the valid confining potentials, in momentum space, range as,
symbolically, $(2\pi)^3K_0\delta^{(3)}(\vec{p})<V(\vec{p})\leqslant(2\pi)^3K_0^3\Delta\delta^{(3)}(\vec{p})$,
that it, from constant to the harmonic oscillator potential, respectively.

For numerical investigation of the mass-gap equation (\ref{mg2}) it is convenient to evaluate analytically
the contribution of the strip $|k-p|<\lambda$ into the integral on the r.h.s.,
\be
I_{\lambda}=\frac{C_FK_0^{\alpha+1}}{2\pi}\frac{\lambda^{2-\alpha}}{2-\alpha}\Gamma(\alpha+1)
\sin\frac{\pi\alpha}{2}[p^2\vp''_p+2p\vp'_p+\sin 2\vp_p],
\label{strip}
\ee
where this integral admits an extra contribution if the width of the strip is chosen different from
$\lambda$. For $\alpha=2$ the dependence of the strip integral (\ref{strip}) on $\lambda$  disappears and
the full mass-gap equation (\ref{diffmge}) is readily reproduced as a consequence of the delta
functional form of the Fourier transform of the harmonic oscillator potential.

\subsection{Numerical analysis}

In this subsection we present the results of the numerical studies of the mass-gap equation (\ref{mg2}) with
$0\leqslant\alpha\leqslant2$. The equation for the harmonic oscillator potential (\ref{diffmge}) is studied in detail in the
literature, so the interested reader can find the details, for example, in refs.~\cite{Orsay,Lisbon}.
In Fig.~1(a) we plot the profile of the nontrivial solutions $\vp_0(p)$ to the mass-gap equation (\ref{mg2})
for several values of $\alpha$. The solution for $\alpha=2$, found in refs.~\cite{Orsay,Lisbon}, is also
depicted for the sake of completeness. In Fig.~1(b) we present the results for the chiral condensate
$\langle \bar
qq\rangle=-\frac{3}{\pi^2}\int_0^{\infty}dp\;p^2\sin\vp_0(p)\equiv-\Lambda_\chi^3$ and for the excess of the
vacuum energy $\Delta{\cal E}_{\rm vac}={\cal E}_{\rm vac}[\vp_0]-{\cal E}_{\rm
vac}[\vp\equiv 0]\equiv-\Lambda_\varepsilon^4$ over the
trivial vacuum as functions of $\alpha$. From Fig.~1(b) one can see that
$\Lambda_\chi\approx\Lambda_{\cal E}$ in the whole range of allowed $\alpha$'s and that their
dependence on the form of the potential is
smooth, solutions for $\alpha$ around unity being quite close to one another (see Fig.~1(a)). Thus we conclude that the
concrete form of the confining potential does not play a crucial role for the physics of chiral symmetry
breaking, resulting only in minor numerical changes. In particular one can see that the qualitative
behaviour of the solution is very stable against deviations of the potential from the purely linear form,
usually adopted in phenomenological models and declared to be confirmed by lattice calculations. Therefore,
at least SBCS and the spectrum of low-lying hadrons will not be strongly affected in case the behaviour of
the confining potential is slightly changed, deviating from linearity, as suggested in \cite{Diakonov}.

\begin{figure}[t]
\begin{tabular}{cc}
\hspace*{-1cm}\epsfig{file=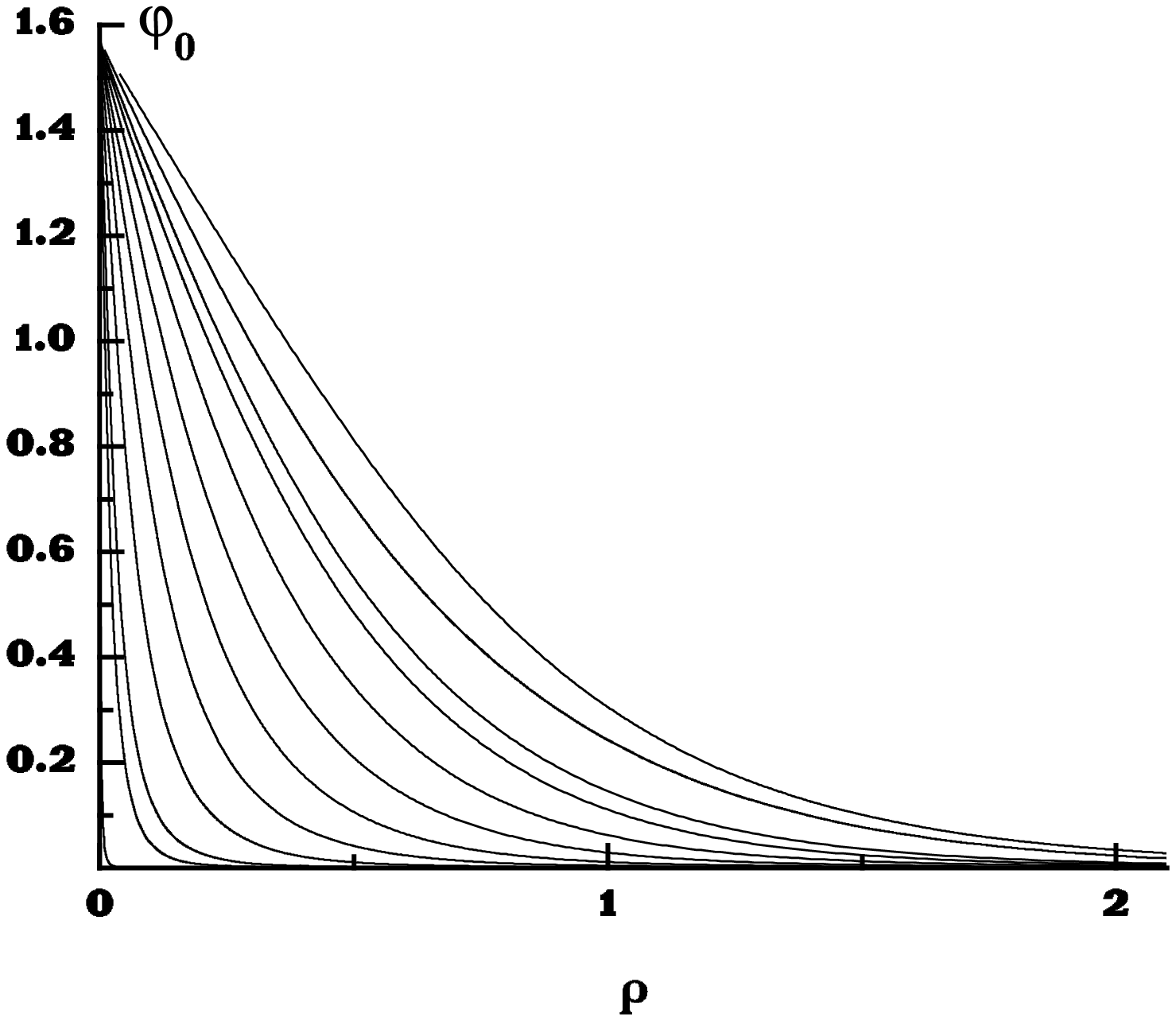,width=9.5cm}&
\hspace*{-2cm}\epsfig{file=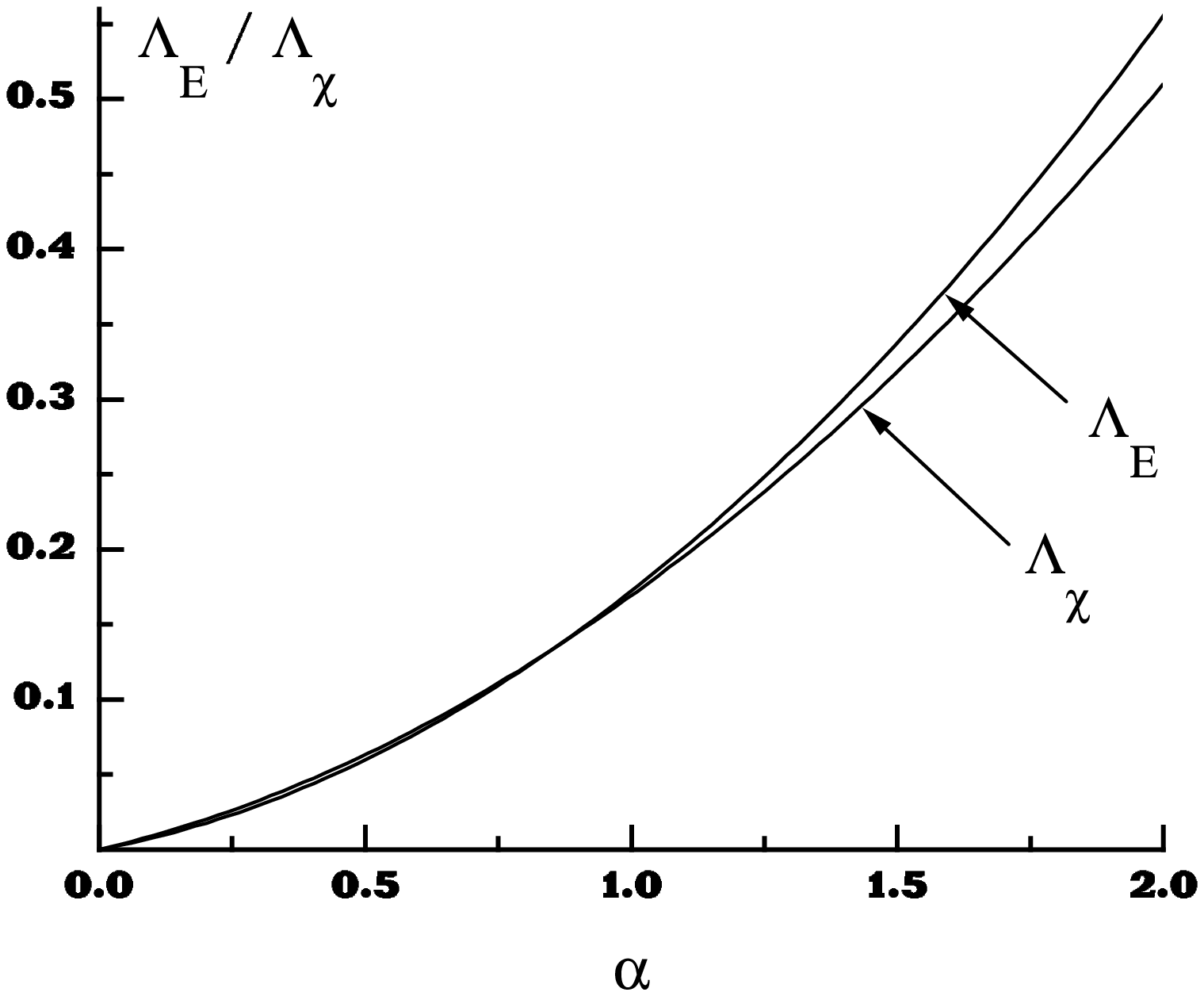,width=9.5cm}\\
\hspace*{-1cm}(a)&\hspace*{-2cm}(b)
\end{tabular}
\caption{The solutions to the mass-gap equation (\ref{mg2}) for $\alpha=0.1$, 0.3, 0.5, 0.7, 0.9, 1.0, 1.1,
1.3, 1.5, 1.7, 1.9, and 2.0 (plot (a)); the curves localised closer to the origin correspond to smaller
$\alpha$'s, and the mass parameters defining the chiral condensate, as
$\langle\bar q q\rangle=-\Lambda_\chi^3$, and the excess of the vacuum energy density over the trivial
vacuum, for two quark flavours and three colours, as $\Delta {\cal E}_{\rm vac}=-\Lambda_{\cal E}^4$,
(plot (b)). All dimensional quantities are given in the units of $K_0$.}
\end{figure}

\section{The replicas}

As argued in a sequence of recent papers \cite{replica1,replica2}, it is possible that \lq\lq$\ldots$
the same ultraviolet behaviour, for instance for the quark propagator, bifurcates to
different solutions when we go to the low-energy domain$\ldots$" in QCD, and such a replica was
discovered for a phenomenology inspired potential. Besides that, the whole infinite tower of excited solutions
for the mass-gap
equation (\ref{diffmge}) for the harmonic oscillator potential was found in
ref.~\cite{Orsay} and also
confirmed in \cite{replica1}. With the
general form of the mass-gap equation (\ref{mgapg}) and (\ref{mg2}) we are in the position to investigate
the problem of replicas existence for various power-like confining potentials.
We find that {\em any} power-like potential $r^{\alpha}$, with the
power $0\leqslant\alpha\leqslant 2$, maintains replicas. In the paper \cite{Orsay} a detailed analysis
was performed for the harmonic oscillator potential mass-gap equation
(\ref{diffmge}) and the existence of an infinite tower of solutions was
proved analytically. We failed to repeat this analysis for the general form of the mass-gap equation
(\ref{mg2}) since, in contrast
to Eq.~(\ref{diffmge}), Eq.~(\ref{mg2}) is integral with the coefficients tuned to provide the overall
convergence of the integral, but leaving no hope to use any expansions under the integral.
Instead let us use an approximate method in order to demonstrate how replica
solutions occur for the mass-gap equation (\ref{mg2}).
To this end we use the parametrisation of the chiral angle through the
effective quark mass and introduce a new function $\psi_p$:
\be
\sin\vp_p=m_pD_p,\quad\cos\vp_p=pD_p,\quad\psi_p=p\sin\vp_p,\quad D_p^{-1}\equiv\sqrt{p^2+m_p^2}.
\label{dfn}
\ee
It is also convenient to use the following integral:
\be
\int_{-\infty}^{\infty}\frac{dk}{2\pi}\frac{\psi_k-\psi_p}{|k-p|^{\alpha+1}}=
-\frac{1}{2\Gamma(\alpha+1)\sin\frac{\pi\alpha}{2}}\int_{-\infty}^{\infty}dx|x|^\alpha\psi_x e^{ipx}.
\ee

Then the mass-gap equation (\ref{mg2}) can be rewritten in a simple physically transparent form:
\be
\left[2E(\hat{p})+C_FK_0^{\alpha+1}|x|^\alpha\right]\psi_x=0,
\label{lin2}
\ee
where the operator $E(\hat{p})$ has the meaning of the quark self-energy and it is given by the
following expression:
\be
E(p)=\frac{1}{D_p}-\frac{C_FK_0^{\alpha+1}}{p^2D_p}\Gamma(\alpha+1)\sin\frac{\pi\alpha}{2}
\int_{-\infty}^{\infty}\frac{dk}{2\pi}\left[\frac{p^2D_p-k^2D_k}{|p-k|^{\alpha+1}}
+\frac{pkD_pD_k}{(\alpha-1)|p-k|^{\alpha-1}}\right].
\label{Epp}
\ee

The even function $m(p)$ takes its maximal value at $p=0$ and then
decreases rapidly. Thus, for momenta larger than some $p_0\sim m(0)$, Eq.~(\ref{lin2})
can be linearised by
putting $D_p^{-1}\approx |p|$ in the expression (\ref{Epp}). Then the integral on the r.h.s. of Eq.~(\ref{Epp}) is
easily evaluated and the self-energy takes the form:
\be
E(p)\approx |p|-\frac{2C_FK_0^{\alpha+1}\Gamma(\alpha+1)\sin\frac{\pi\alpha}{2}}{\pi\alpha(2-\alpha)|p|^\alpha}.
\label{Epp3}
\ee

Notice that for {\em any confining potential}, $\alpha\geqslant 0$, the second term on the r.h.s. of the
expression (\ref{Epp3}) dominates in the low-momentum region bringing a large negative contribution to the
quark self-energy (the corresponding term becomes logarithmic, $\sim K_0\ln\frac{K_0}{|p|}$, for the
potential (\ref{ln})). This general feature of the quark self-energy in the confining potential has been
discussed in literature not once (see, for example, \cite{Orsay,Lisbon}, or the papers \cite{2d}
where the case of QCD$_2$ is discussed in detail) and it is known to play a crucial role for the properties
of the theory. Besides that, replicas exist in the theory also due to this feature of the quark self-energy
$E(p)$. In order to demonstrate this let us write the linearised mass-gap equation (\ref{lin2}) in the form of
a Schr{\" o}dinger-type equation in momentum space,
\be
\left[C_FK_0^{\alpha+1}|\hat{x}|^\alpha+2|p|
-\frac{4C_FK_0^{\alpha+1}\Gamma(\alpha)\sin\frac{\pi\alpha}{2}}{\pi(2-\alpha)|p|^\alpha}\right]\psi_p=
\varepsilon\psi_p,
\label{lin3}
\ee
and notice that the linearisation suppresses the first, positive, term on the r.h.s. of Eq.~(\ref{Epp}) and
enhances the second, negative, term. As a result, we are interested in the {\em odd} (see the definition
(\ref{dfn})) eigenstates of the
linear equation (\ref{lin3}) with {\em negative} eigenvalues. Each such state indicates a solution of the
full nonlinear mass-gap equation (\ref{mg2}) and can be used as the starting anzatz for the iterative
numerical method to solve the latter. Therefore, in order to find the number of solutions to the
mass-gap equation (\ref{mg2}), it is sufficient to count the odd eigenstates, with negative eigenvalues, of
the linear equation
(\ref{lin3}). If the Bohr-Sommerfeld quantisation procedure is applied directly to Eq.~(\ref{lin3}), then the
quasiclassical integral, $I_{WKB}=\int_{p_{\rm min}}^{p_{\rm max}}x(p)dp$, diverges logarithmically at $p=0$, where
the chiral angle is not small anymore and the approximation $D_p^{-1}\approx |p|$ obviously fails.
In this region, the sharp behaviour of the self-energy is smeared by the effective quark mass,
$\mu\sim m_p(p\to 0)$, which plays the role of the effective regulator, so that we modify Eq.~(\ref{lin3})
accordingly:
\be
\left[C_FK_0^{\alpha+1}|\hat{x}|^\alpha+2\sqrt{p^2+\mu^2}
-\frac{4C_FK_0^{\alpha+1}\Gamma(\alpha)\sin\frac{\pi\alpha}{2}}{\pi(2-\alpha)(p^2+\mu^2)^{\alpha/2}}\right]
\psi_p=\varepsilon\psi_p.
\label{lin33}
\ee

The spectrum of eigenstates of Eq.~(\ref{lin33}) starts, for $n=0$, at the bottom of the deep well described
by the effective potential $V(p)=2E(p)$, $\varepsilon_0\approx V(p=0)\sim -K_0^{1+\alpha}/\mu^\alpha$, and, for some $n_{\rm max}$, reaches
zero from below. Then the spectrum continues for positive eigenvalues up to infinity. Since the last negative
eigenenergy $\varepsilon_{n_{\rm max}}$ is small, then we expand the quasiclassical integral accordingly and find:
\be
\varepsilon_n\mathop{\approx}\limits_{n\approx n_{\rm max}}
K_0\left(\frac{C_F\Gamma\left(\frac12\right)\Gamma\left(\frac{4-\alpha}{2}\right)}{
\Gamma\left(\frac{1+\alpha}{2}\right)}\right)^{\frac{1}{1+\alpha}}
\frac{\Gamma\left(\frac{1}{\alpha(1+\alpha)}\right)}{\Gamma\left(\frac{1}{\alpha}\right)
\Gamma\left(\frac{\alpha}{1+\alpha}\right)}\left[\pi n\left(
\frac{\sqrt{\pi}\Gamma\left(\frac{4-\alpha}{2}\right)}{2^\alpha\Gamma\left(\frac{1+\alpha}{2}\right)}
\right)^{\frac{1}{\alpha}}-\ln\frac{K_0}{\mu}\right],
\label{nnn}
\ee
where we expressed $\sin\frac{\pi\alpha}{2}$ through the Euler Gamma functions as
\be
\sin\frac{\pi\alpha}{2}=2^{\alpha-1}\frac{\Gamma\left(\frac{1}{2}\right)\Gamma\left(\frac{1+\alpha}{2}\right)}
{\Gamma(\alpha)\Gamma\left(\frac{2-\alpha}{2}\right)}.
\label{sin}
\ee

Notice that in actuality the regulator $\mu$ itself is not a constant, but it rapidly decreases with $n$, and it is such
that all eigenvalues of Eq.~(\ref{lin33}) vanish for all $n$'s. Then the corresponding eigenfunctions
$\psi_n$ provide, according to the definition (\ref{dfn}), the solutions to the exact mass-gap equation
(\ref{mg2}). From Eq.~(\ref{nnn}) one can easily see that
\be
\mu_n(\alpha)=K_0\exp{\left(-C_\alpha \pi n\right)},\quad
C_\alpha=\left[\frac{\sqrt{\pi}\Gamma\left(\frac{4-\alpha}{2}\right)}{2^\alpha\Gamma
\left(\frac{1+\alpha}{2}\right)}\right]^{1/\alpha}.
\label{Calpha}
\ee

Using a more accurate expansion in
Eq.~(\ref{nnn}), one can find the correction $\delta_\alpha$ to the leading regime (\ref{Calpha}),
$\mu_n(\alpha)=K_0\exp{\left(-C_\alpha(\pi n+\delta_\alpha)\right)}$, as it 
was done in \cite{Orsay}. Unlike the leading logarithmic term, this correction 
is sensitive to the concrete form of the regularization of the potential 
$V(p)$, and we do not give it here.

From the formula (\ref{Calpha}) we conclude that, for any $\alpha$, the mass-gap equation
(\ref{mg2}) supports an infinite number of solutions which, in the low-momentum region $(p\ll K_0)$, behave as
\be
\vp_n(p\sim 0)=\frac{\pi}{2}-\frac{p}{K_0}e^{C_\alpha(\pi n+\delta_\alpha)}+\ldots,
\label{slor}
\ee
then oscillate $n$ times and, finally, approach zero for the infinite momentum. The constant
$C_\alpha$ is monotonously decreasing function of $\alpha$ in the whole interval
$0\leqslant\alpha\leqslant 2$ and it varies from
$C_0=\frac12\exp[\frac12(\gamma-1-\psi(\frac12))]\approx 1.08$
to $C_2=\frac{1}{\sqrt{2}}\approx 0.71$.
Numerically, the same behaviour of the solutions to the mass-gap equation (\ref{mg2}) is observed ---
for smaller $\alpha$'s and higher $n$'s the chiral angle becomes steeper at the origin, in accordance with the obvious
identification $\vp'_{n|p=0}=-1/\mu_n\propto\exp{(C_\alpha [\pi n+\delta_\alpha])}$ (see Figs.~1(a), 2).

The formula (\ref{Calpha}) is approximate since the dependence on $\mu$ in Eq.~(\ref{lin33})
reproduces only the gross features of the quark self-energy, whereas the exact dependence of the self-energy on
the chiral angle is, in turn, a consequence of the mass-gap equation, and the problem becomes self-consistent.
To estimate the accuracy of the formula (\ref{Calpha}) we continue it to $\alpha=2$ and
compare our result with the result found in \cite{Orsay}:
$$
C_2=\frac{1}{\sqrt{2}}\approx 0.71,\quad C_2({\rm ref.}~[3])=\frac{2}{\sqrt{7}}\approx 0.76,
$$
that is, the error is about $6\div 7$\%.

As an indirect confirmation of the conclusion made above let us mention that no critical phenomena are observed
in the mass-gap equation for any $\alpha$ as well as in the limit $\alpha\to 2$, so that all properties of the
mass-gap equation, including the infinite number of solutions, may be continued from $\alpha=2$ to smaller
$\alpha$'s.
\begin{figure}[t]
\epsfig{file=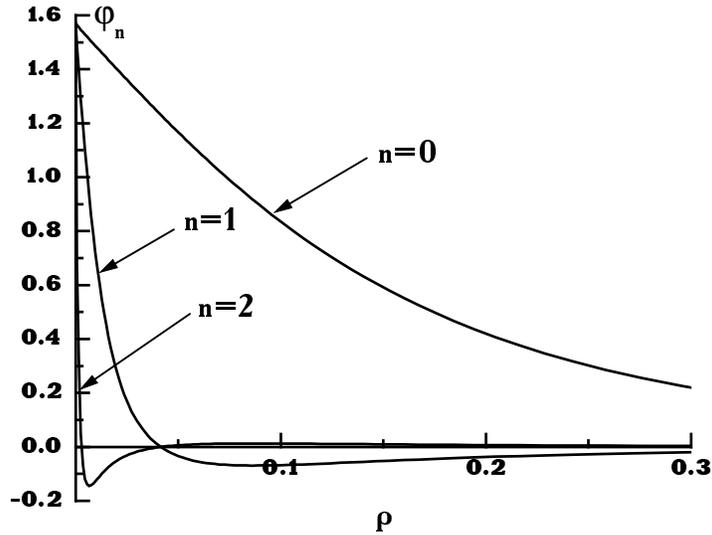,width=12cm}
\caption{The three first solutions to the mass-gap equation (\ref{lmge}) corresponding to the ground-state
BCS vacuum ($n=0$) and to the first two replica states ($n=1,2$). The momentum $p$ is given in the units of
$K_0$.}
\end{figure}

In Fig.~2 we give the profiles of the ground-state, as well as of the first two replica solutions
for the linear confinement, $\alpha=1$. Solutions with larger number
of nodes are also available for numerical study, but the method has to be very precise since each
new solution possesses oscillations squeezed to zero and seen only with the help of methods with a
sufficiently high resolution. According to the formula (\ref{slor}), the behaviour of these
solutions at the origin can be approximated as
\be
\vp_n(p\sim 0)=\frac{\pi}{2}-\frac{p}{K_0}\exp{\left[\frac{\pi}{4} (\pi n+\delta_1)\right]}+\ldots,
\label{slor2}
\ee
so that $\mu_{n+1}(1)/\mu_n(1)\approx\exp{(\frac{\pi^2}{4})}\sim 10$, that is, the parameter of SBCS
decreases about ten times for each next replica.

In ref.~\cite{replica2} a method is proposed which allows one to take
the vacuum replicas into account in quark models. The generalisation of this approach
to the case of many replicas is straightforward and assumes the summation over contributions of all replicas.
On the other hand, since the mass parameter of the solution, for example, $\mu_n$, decreases fast
(approximately ten times for the linear confinement) for each next solution, then one has a sort of
perturbative series with the well defined
convergence parameter. Therefore it may be sufficient to consider only the first
replica neglecting the contribution of higher replicas, starting from the second one, which are hard to
distinguish numerically from the trivial solution $\vp_p\equiv 0$. To have a good phenomenological
description of quarkonia one should supply the purely confining potential by the short-range Coulomb
interaction and, possibly, by a constant term in order to fit for the right value 
of the chiral condensate (an attampt to evaluate this constant from first principles was undertaken 
in \cite{sener}).
Such a potential was considered in \cite{replica1,replica2} and the solutions for the ground-state BCS
vacuum as well as for one vacuum replica were found numerically.

\section{Conclusions}

In this paper we complete the study of the power-like potentials $K_0^{\alpha+1}r^\alpha$
from the point of view of SBCS and the number of nontrivial solutions of the mass-gap equation. We
use the constructive method to prove the chirally symmetric vacuum instability for such
potentials, solving the mass-gap equation explicitly and calculating the vacuum energy for
the corresponding solution. We establish the region of allowed powers of $\alpha$ for such confining
potentials which lead to a convergent mass-gap equation and a finite excess of the vacuum
energy density over the trivial solution with unbroken chiral symmetry. Thus, using simple
qualitative arguments, we demonstrate that for $0\leqslant\alpha<d$, $d$ being the number of spatial
dimensions, at least two, chirally symmetric and nonsymmetric, solutions should
exist, whereas for $\alpha\geqslant d$ the trivial solution possesses an infinite energy density and
disappears. In the meantime, the restriction for the parameter $\alpha$ exists,
$0\leqslant\alpha\leqslant 2$, which comes from the fact that the corresponding mass-gap equation should be
convergent. We find numerically solutions to the mass-gap equation for various values of $\alpha$
from the allowed region and, for the found solutions, evaluate the vacuum energy density
and the chiral condensate which are given by the same scale, $\Lambda_\varepsilon\sim\Lambda_\chi$, and
turn out to be slow functions of the parameter $\alpha$.

We address the question of existence of the second, third, and higher chirally nonsymmetric solutions of the
mass-gap equation for power-like confining potentials and find that any
potential with $0\leqslant\alpha\leqslant 2$ supports such solutions --- the replicas --- which is rather the
rule for confining potentials and comes from the peculiar behaviour of the quark self-energy in
the infrared domain. We find that the number of such replicas is infinite for any $\alpha$ and estimate the
slope of the solutions to the mass-gap equation at the origin.

Thus we dare predict the existence of replicas regardless of the explicit form of the confinement and of the
details of the model used in calculations. In real QCD, with light quark flavours and the quark-quark
potential flattening at large distances due to the effect of the QCD string breaking 
(such an effect can be taken into account in the interquark potential through a
coordinate dependence of the effective string tension \cite{flat}), the number of replicas
is expected to be finite. Indeed, the distance at which the string starts to break, $L$, will play the role of
the infrared regulator in the formula similar to (\ref{lin33}), instead of $\mu$. Therefore, as follows from
Eq.~(\ref{nnn}),
$n_{\rm max}\sim\ln(K_0L)\sim 1$, where we consider $L\sim K_0^{-1}$. In other words, confinement becomes less
\lq\lq binding" due to string breaking and the corresponding mass-gap equation supports fewer solutions ---
\lq\lq bound states". A more detailed analysis of the QCD inspired interaction, including the proper string
dynamics, from the point of view of replicas is in progress now and will be subject for future publications.

\begin{acknowledgments}
The authors are grateful to A.~A.~Abrikosov Jr. for fruitful discussions, as well as to
Yu.~S.~Kalashnikova and J.~E.~Ribeiro for reading the manuscript and valuable comments. One of the
authors (A.V.N.) would like to thank the
staff of the Centro de F\'\i sica das Interac\c c\~oes Fundamentais (CFIF-IST) for
cordial hospitality during his stay in Lisbon where this work was originated
and to acknowledge the financial support of INTAS grants OPEN 2000-110 and
YSF 2002-49, as well as the grant NS-1774.2003.2.
\end{acknowledgments}

\end{document}